\documentclass[fleqn,usenatbib]{mnras}

\usepackage[usenames,dvipsnames]{xcolor}
\usepackage{amsmath}
\usepackage{array}
\usepackage{xfrac}
\newcommand{\HI}{{\sc H\,i}}
\usepackage{multirow}

\usepackage[T1]{fontenc}
\usepackage{ae,aecompl}
\hypersetup{draft}

\title[Milky Way and Andromeda Analogs]{Are the Milky Way and Andromeda unusual? A comparison with Milky Way and Andromeda Analogs}

\author[Boardman et al.]{
N.~Boardman$^{1}$\thanks{E-mail: nick.boardman@astro.utah.edu},
G.~Zasowski$^{1}$,
J.~A.~Newman$^{2}$,
B.~Andrews$^{2}$, 
C.~Fielder$^{2}$,\newauthor
M.~Bershady$^{3,4,5}$, 
J.~Brinkmann$^{6}$,
N.~Drory$^{7}$,
D.~Krishnarao$^{3}$,
R.~R.~Lane$^{8,9}$,\newauthor
T.~Mackereth$^{10}$,
K.~Masters$^{11}$,
G.~S.~Stringfellow$^{12}$
\\
$^{1}$Department of Physics \& Astronomy, University of Utah, Salt Lake City, UT, 84112, USA\\
$^{2}$Department of Physics \& Astronomy and PITT PACC, University of Pittsburgh, Pittsburgh, PA 15260, USA\\
$^{3}$Department of Astronomy, University of Wisconsin-Madison, 475N. Charter St., Madison WI 53703, USA\\
$^{4}$South African Astronomical Observatory, P.O. Box 9, Observatory 7935, Cape Town, South Africa\\
$^{5}$Department of Astronomy, University of Cape Town, Private Bag X3, Rondebosch 7701, South Africa\\
$^{6}$Apache Point Observatory, P.O. Box 59, Sunspot, NM 88349, USA\\
$^{7}$McDonald Observatory, The University of Texas at Austin, 1 University Station, Austin, TX 78712, USA\\
$^{8}$Pontificia Universidad Cat\'  olica de Chile, Instituto de Astrofısica, Av. Vicuna Mackenna 4860, 782-0436 Macul, Santiago, Chile\\
$^{9}$"Instituto de Astronom\'ia y Ciencias Planetarias, Universidad de Atacama, Copayapu 485, Copiap\'o, Chile"\\
$^{10}$School of Physics \& Astronomy, University of Birmingham, Birmingham, B15 2TT, UK\\
$^{11}$Department of Physics and Astronomy, Haverford College, 370 Lancaster Ave, Haverford, PA 19041\\
$^{12}$Center for Astrophysics and Space Astronomy, Department of Astrophysical and Planetary Sciences, University of Colorado, 389 UCB, Boulder, CO 80309-0389, USA
}

\date{Accepted XXX. Received YYY; in original form ZZZ}
\pubyear{2019}
\begin{document} 
\label{firstpage}
\pagerange{\pageref{firstpage}--\pageref{lastpage}}
\maketitle

\begin{abstract}

Our Milky Way provides a unique test case for galaxy evolution models, thanks to our privileged position within the Milky Way's disc. This position also complicates comparisons between the Milky Way and external galaxies, due to our inability to observe the Milky Way from an external point of view. Milky Way analog galaxies offer us a chance to bridge this divide by providing the external perspective that we otherwise lack. However, over-precise definitions of ``analog'' yield little-to-no galaxies, so it is vital to understand which selection criteria produce the most meaningful analog samples. To address this, we compare the properties of complementary samples of Milky Way analogs selected using different criteria. We find the Milky Way to be within 1$\sigma$ of its analogs in terms of star-formation rate and bulge-to-total ratio in most cases, but we find larger offsets between the Milky Way and its analogs in terms of disc scale length; this suggests that scale length must be included in analog selections in addition to other criteria if the most accurate analogs are to be selected. We also apply our methodology to the neighbouring Andromeda galaxy. We find analogs selected on the basis of strong morphological features to display much higher star-formation rates than Andromeda, and we also find analogs selected on Andromeda's star-formation rate to over-predict Andromeda's bulge extent. This suggests both structure and star-formation rate should be considered when selecting the most stringent Andromeda analogs.

\end{abstract}


\section{Introduction}

The question of how galaxies form and evolve remains a problem of significant interest in extragalactic astrophysics. The last two decades have seen major progress on this topic, thanks in particular to the census of local Universe galaxies undertaken by the Sloan Digital Sky Survey \citep[SDSS;][]{york2000}. SDSS data conclusively demonstrated a bimodality in terms of both galaxies' integrated colors \citep{strateva2001} and integrated magnitudes \citep{baldry2004}, with the majority of galaxies being either star-forming ``blue cloud" galaxies or quiescent ``red sequence" galaxies. At the same time, a minority of galaxies can be seen to occupy the intermediate ``green valley" when plotted on color-magnitude diagrams. It is commonly accepted that the green valley is occupied by galaxies in the process of transitioning from star-forming to quiescent. However, a large number of precise evolution pathways are needed to fully explain the range of observed galaxy properties such as integrated color, star-formation rate and morphology \citep[e.g.][]{bell2004,faber2007, schawinski2014,smethurst2015}.

Our own Milky Way (MW) remains a key source of insight into both the structures of galaxies and the physics of galaxy evolution on small scales. Traditionally, the MW has been understood to contain both a  younger ``thin disc" and an older alpha-enhanced ``thick disc" \citep{yoshii1982,gilmore1983,chiappini1997,bensby2003,haywood2013,xiang2017,wu2019}, which produces an observed bimodality in alpha abundance ratios at intermediate MW stellar metallicities \citep[e.g.][]{fuhrmann1998,fuhrmann2011,anders2014,nidever2014,mikolaitis2014,reciobianco2014,hayden2015}; such a bimodality is observed in both the MW's disc \citep[e.g.][]{hayden2015} and bulge \citep{rojasarriagada2019,queiroz2019}. The MW's inner region is dominated by a cylindrical-rotating boxy-peanut ``pseudobulge" \citep[e.g.][]{dwek1995,mcwilliam2010} along with a stellar bar \citep[e.g.][]{hammersley1994, weiland1994,wegg2013}. In addition, the MW may also contain a small ``classical" bulge component made up of old stars \citep[e.g.][]{dekany2013,barbuy2018}.

The last decade has proven particularly fruitful in regards to our understanding of the MW, with a number of spectroscopic surveys observing stars over large regions of the MW's stellar disc. These surveys include RAVE \citep{steinmetz2006}, LEGUE \citep{deng2012}, LAMOST \citep{Cui_2012_lamost}, the Gaia-ESO survey \citep{gilmore2012}, GALAH \citep{desilva2015}, and APOGEE \citep{majewski2017}. Such data sets have challenged the traditional two-population view of the MW's stars, with the MW instead appearing to consist of a broad continuum of chemo-dynamical stellar subpopulations \citep[e.g.][]{bovy2012a, mackereth2017, buder2019}. Furthermore, it has become clear that the geometrical thin and thick discs of the MW are \textit{not} the same structures as the chemical thin and thick discs \citep[e.g.][]{minchev2015,martig2016}.

However, the MW's position within the galactic population remains poorly understood, which complicates the interpretation of MW results in the wider extragalactic context. This situation results from our position within the MW's disc, from which large regions of the MW remain difficult to observe. Dust extinction preferentially reddens stars away from the solar region, limiting the number of stars that can be observed particularly at UV and optical wavelengths \citep[e.g.][]{schlegel1998,schlafly2011,queiroz2019}; this especially complicates the analysis of stars beyond the central stellar bulge and bar. It is therefore difficult to determine integrated quantities for the MW such as optical color \citep[e.g.][]{mutch2011}, which can be straightforwardly calculated for nearby galaxies.

It thus remains unclear how common the MW's properties are within the wider galaxy population. A particular open question concerns the size of the MW's disc component, which has repeatedly been argued to be comparatively low \citep[e.g.][]{bh2016,licquia2016a}. The MW disc appears to be more compact than the majority of galaxies of MW-like mass \citep[e.g.][hereafter LN16]{bovy2013,licquia2016} and appears to be deficient by around 1$\sigma$ with respect to the MW's circular velocity \citep{hammer2007}, whereas the scale length of M31 appears far more usual \citep{hammer2007}. The MW displays properties in good consistency with other galaxies once its disc's compactness is taken into account, meanwhile \citep{bovy2013, boardman2020}.

Milky Way Analogs (MWAs) provide an ideal opportunity to bridge the gap between Galactic and extragalactic observations. MWAs allow one to estimate global properties of the MW that cannot be easily or directly estimated for the MW, and have enabled tight estimates of the MW's magnitude and color \citep{L15}. MWAs also allow comparison of the MW to its immediate peers \citep[e.g.][]{frasermckelvie2019b,boardman2020,dk2020}. Thus, MWAs are a powerful tool for better understanding the Milky Way in the extragalactic context. 

However, there is no one definition for what makes a galaxy a MWA, and overly-strict definitions of ``analog" can produce negligible or even flat-out nonexistent samples of MWA galaxies. \citet{frasermckelvie2019b}, for instance, find just 176 MWAs from selecting on stelar mass, bulge-to-total ratio (B/T) and morphology. \citet{boardman2020}, meanwhile, find not a single MWA in the SDSS-IV \citep{blanton2017} MaNGA survey \citep{bundy2015} when attempting to select on a combination of stellar mass ($M_*$), star-formation rate (SFR),
B/T and disc scale length ($R_d$). It is thus critical to assess the impacts of different selection criteria, in order to understand how to best select constraining samples of MWAs.

M31 Analogs, hereafter M31As, can provide us with additional insight. M31As allow us to assess Andromeda's position amongst its peers in a similar manner to what MWAs enable for the MW. The MW and M31 are the two nearest massive disc galaxies to us, and can both be studied in much greater depth than other such galaxies; thus, it is crucial to understand how both the MW and M31 relate to the wider extragalactic population.

We experiment here with a number of complementary selection criteria, aimed at selecting various samples of ``analogs" and then comparing their ranges of properties to our knowledge of the MW. We perform an equivalent analysis on M31A galaxies, selecting M31 analogs through multiple means and then comparing with M31. 

This paper is organised as follows. In \autoref{sample} we describe the methodology behind our various sample selections, and then in \autoref{samplecompare} we present our results in terms of the samples' properties and their comparison to the MW and to M31. We discuss our findings and conclude in \autoref{discussion}.

\section{Sample and Data}\label{sample}

\begin{table*}
\begin{center}
\begin{tabular}{l|c|c|c|c}
Analog Sample & SFR ($M_\odot$~yr$^{-1}$) & B/T & $R_d$ (kpc) & Galaxy Zoo vote fraction \\
\hline
\hline
MW star-formation analogs & $1.46-1.84$ & --- & --- & --- \\
MW bulge analogs & --- & $0.13-0.19$ & --- & --- \\
MW scale analogs & --- & --- & $2.51-2.93$ & --- \\
MW morphological analogs & --- & --- & --- & spiral $> 0.8$, bar $>0.8$, $N_{\rm bar, spiral} \geq 20$ \\
 &  &  & &  smooth $\leq 0.57$, edgeon $\leq 0.285$ \\
\end{tabular}
\end{center}
\caption{MWA sample definitions, in addition to $M_* = 4.6-7.2 \times 10^{10} M_\odot$, $z \leq 0.06$, $P_{pS} \leq 0.32$ and $b/a \geq 0.6$.}
\label{testtable}
\end{table*}

In \autoref{samplesources} we discuss the acquisition of all parameters being considered in our selections as well as in our analysis. We discuss our MWA selections in \autoref{samplemwas} and our M31A selections in \autoref{samplem31as}

\subsection{Source catalogs}\label{samplesources}

We obtain stellar masses, SFRs, B/T ratios, disc scale lengths, multi-band magnitudes, redshifts, and quantitative galaxy morphologies by cross-referencing a number of published catalogs. We consider in our subsequent analysis only galaxies that are present in all catalogs. 

We obtain \textbf{total stellar masses and current global star-formation rates} from the GALEX-SDSS-WISE Legacy catalog \citep[GSWLC;][]{salim2016}, employing the GSWLC-2X catalog \citep{salim2018}. GSWLC-2X contains stellar masses and SFRs for 659229 galaxies, selected by cross-referencing the spectroscopic SDSS Data Release 10 \citep[DR10;][]{ahn2014} sample with the ultraviolet Galaxy Evolution Explorer \citep[GALEX;][]{martin2005} sample. Masses and SFRs are derived through spectral energy distribution (SED) fits performed using the Code Investigating GALaxy Emission \citep[CIGALE;][]{noll2009,boquien2019}, which uses WISE \citep{wright2010} IR luminosity as a constraint on UV-optical fits to combined GALEX-SDSS photometry. The GSWLC masses and SFRs assume a \citet{chabrier2003} initial mass function (IMF); we converted these to a \citet{kroupa2001} IMF by multiplying by a factor of 1.06 \citep{elbaz2007,salim2007,zahid2012}.

We obtain \textbf{light-weighted r-band B/T and $R_d$ values} from the catalog of \citet{simard2011}, who perform bulge-disc decompositions of 1,123,718 galaxies from the SDSS Data Release 7 \citep[DR7;][]{abazajian2009}. Specifically, we employed the two-component bulge+disc fits from that paper in which the bulge S\'ersic index $n_b$ was treated as a free parameter. \citet{simard2011} perform their fits simultaneously to g-band and r-band SDSS galaxy photometry; structural parameters including $n_b$ and $R_d$ are fixed to be identical in both bands, whereas the amplitude of individual components (and hence B/T) are allowed to vary between bands. We also obtain from this catalog the $P_{pS}$  (``probability of pure S\'ersic") parameter for each galaxy. $P_{pS}$ denotes the \textit{F}-test probability of a bulge+disc model \textit{not} being required to fit a given galaxy, as opposed to a pure S\'ersic model, and parametrises the goodness-of-fit improvement achieved by fitting a bulge+disc component over a single S\'ersic component.  

We obtain \textbf{redshifts and absolute magnitudes} from version 1.0.1 of the NASA-Sloan Atlas\footnote{ http://www.nsatlas.com} (NSA) catalog \citep{blanton2011}, which re-reduces the data in SDSS Data Release 8 \citep[DR8;][]{aihara2011}. The NSA absolute magnitudes are drawn from a combination of GALEX and SDSS photometry, and are provided over seven bands (FNugriz) overall; we use the elliptical Petrosian set of values throughout our analysis. For redshifts, we use values obtained from the distance estimates of \citet{willick1997}

Finally, we obtain measurements of \textbf{galaxy morphologies} from the Galaxy Zoo 2 catalog \citep[GZ2;][]{willett2013}, employing  user-weighted vote fractions along with the redshift-debiased fractions described in \citet{hart2016}. We obtained information on the presence of bar and spiral features along with information concerning the galaxies' inclinations. The Galaxy Zoo vote-based method allows quantitative measures of morphology as well as providing quantitative confidence levels in those morphologies, and has repeatedly been shown to be an excellent means of detecting spiral arms and bars \citep[e.g.][]{hart2017,kruk2018}. The specific parameters we extracted from the catalog were ``t01\_smooth\_or\_features\_a01\_smooth\_weighted\_fraction", ``t02\_edgeon\_a04\_yes\_weighted\_fraction", ``t03\_bar\_a06\_bar\_debiased", ``t04\_spiral\_a08\_spiral\_debiased", ``t03\_bar\_a06\_bar\_count" and ``t04\_spiral\_a08\_spiral\_count"; for the remainder of this paper, we will refer to these parameters respectively as \textit{smooth}, \textit{edgeon}, \textit{bar}, \textit{spiral}, $N_{bar}$ and $N_{spiral}$. We will also refer to a parameter $N_{bar,spiral}$, describing the lower value out of $N_{bar}$ and $N_{spiral}$ for a given galaxy.

We obtain a total of 149585 galaxies from this cross-referencing procedure. Following \citet{simard2011}, we discount all galaxies for which $P_{pS} > 0.32$, as such galaxies are less likely to be true bulge+disc systems and so are more likely to yield spurious bulge-disc decompositions; this yields 82724 galaxies. We then remove all galaxies with elliptical Petrosian axis ratios (b/a) below 0.6, to avoid considering discy galaxies with edge-on viewing angles and so strong internal dust extinction \citep[e.g.][]{L15}, leaving 62735 galaxies; we discuss this cut further in the following subsection. We further remove 1337 galaxies for which $R_d < 1$ and $B/T > 0.8$, in order to eliminate galaxies with unreliable bulge+disc fits that were not eliminated by the previous cut. This produces a final parent sample of 61398 galaxies, from which we select MWA and M31A samples as described in the following subsections.

We employ \textit{volume-limited} analog samples throughout our analysis. Essentially, we wish to ensure that we do not miss fainter MWAs/M31As due to the magnitude limits of employed galaxy catalogs, allowing unbiased comparisons with the MW and M31. We achieve this by restricting analog samples to a given maximum redshift, as described in the next two subsections for MWAs and M31As respectively. 

\subsection{Milky Way analog sample selections}\label{samplemwas}

We select a series of MWA samples based on our knowledge of the MW. Each sample is based on a different definition of ``analog", as described in the remainder of this section and summarised in \autoref{testtable}. 

Our MWA selections are largely based on the MW parameter values reported in \citet{licquia2015} (hereafter LN15) and LN16. These two works combine a wide variety of literature measurements in order to obtain constraints on the MW's $M_*$, SFR, B/T and $R_d$ values. LN15 determine the MW SFR by performing a Heiarchical Bayesian analysis on a range of previous SFR measurements retrieved from Table 1 of \citet{chomiuk2011}. LN16 similarly obtain a MW $R_d$ value by combining numerous individual measurements \citep[e.g.][]{kent1991,rm1991,chen1999,benjamin2005,chang2011,mao2015}, with the value effectively being an estimate for the MW's thin disc. $M_*$ and $B/T$ values are determined by combining stellar bulge and/or bar mass values from the literature \citep[e.g.][]{kent1992,dwek1995,widrow2008} and then using these in conjuction with Monte Carlo simulations of an exponential disc model based on \citet{bovy2013}. LN15 employ an $R_d$ value of $2.15 \pm 0.14$ for this process, from \citet{bovy2013}, whereas LN16 employ their own $R_d$ estimate.

The galactic parameters we consider are likely to vary over a variety of different timescales. In particular, the SFR is expected to fluctuate relatively rapidly over a galaxy's lifetime, whereas a galaxy's mass and stucture will change far more gradually. In general, the key assumption behind MWA selections is that the MW \textit{should not be unusual} amongst a sample of its chosen analogs \citep[e.g.][]{L15,boardman2020}, regardless of the specific parameters employed in selections. Thus, it is worthwhile to consider both long-lived and shorter-lived parameters when selecting MWA samples.

\begin{figure}
\begin{center}
	\includegraphics[trim = 1cm 3cm 10cm 2cm,scale=0.9
	,clip]{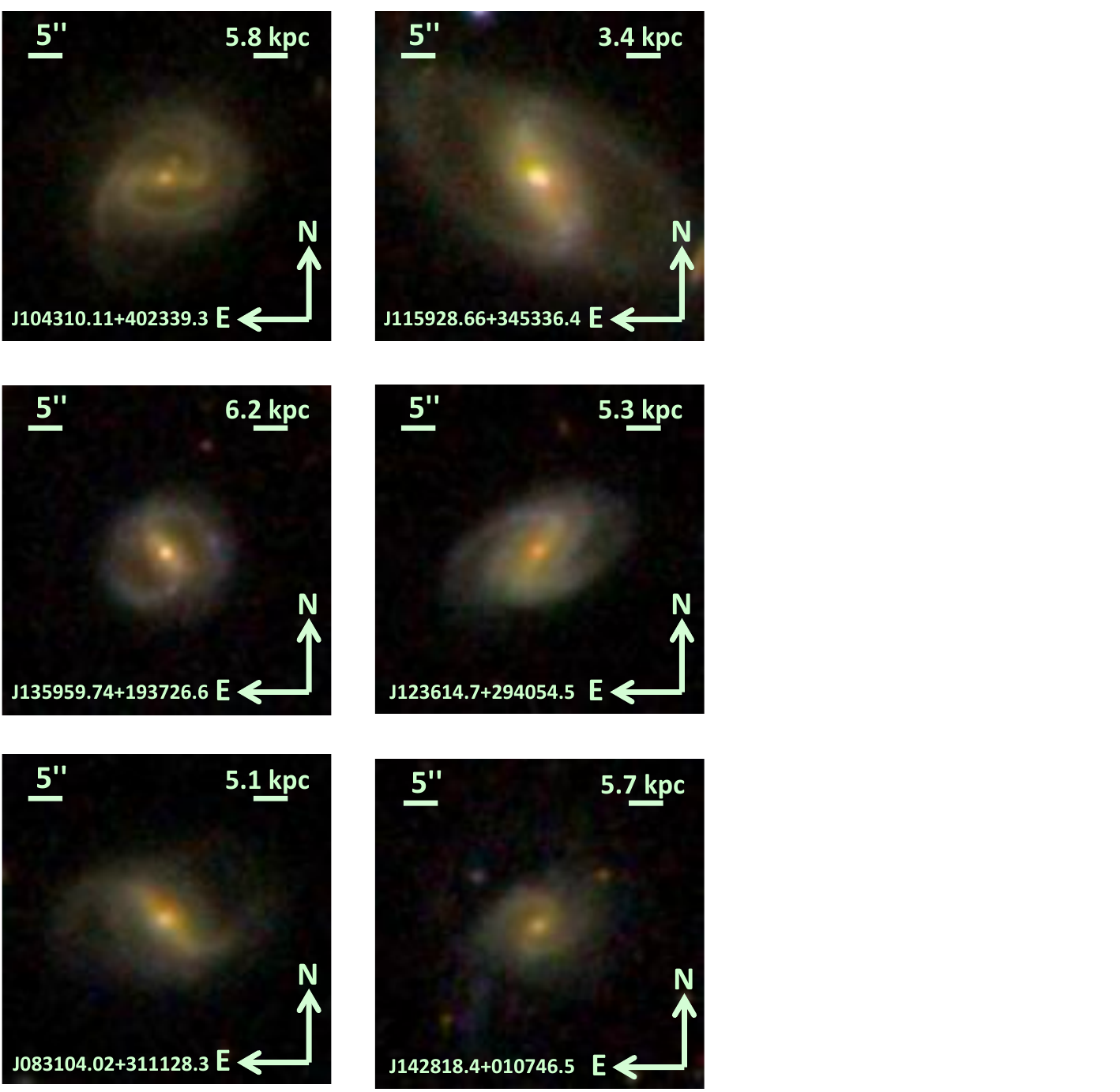}
	\caption{Example morphological MWAs, selected on mass and on Galaxy Zoo morphology votes as discussed in the text.} 
	\label{morphmwaexamples}
	\end{center}
\end{figure}

We select \textbf{``star-formation MWAs"} (hereafter \textbf{``SF MWAs"}) based on stellar masses and star-formation rates. This is a natural MWA definition, as both parameters are known to be strongly correlated with galaxies' global photometric properties \citep{L15}. We select galaxies with stellar masses between $4.6 \times 10^{10} M_\odot$ and $7.2 \times 10^{10} M_\odot$, based on the 1-$\sigma$ confidence intervals reported in LN15. We then further limit the sample to those galaxies with SFR values between 1.46 and 1.84 $M_\odot /\textrm{yr}$, based on the 1-$\sigma$ interval found in LN16.

We select \textbf{``bulge MWAs"} by cutting on stellar masses and bulge-to-total ratios. This is another natural choice for performing MWA selections, due to the known connection between the growth of a galaxy's bulge with a galaxy's particular evolution history \citep[e.g.][]{cappellari2016,belfiore2017a,saha2018}. We perform the same mass cut as before, and we select galaxies with B/T values between 0.13 and 0.19 based on the LN16 1-$\sigma$ intervals. 

\begin{figure}
\begin{center}
	\includegraphics[trim = 3cm 10cm 3cm 7cm,scale=0.4]{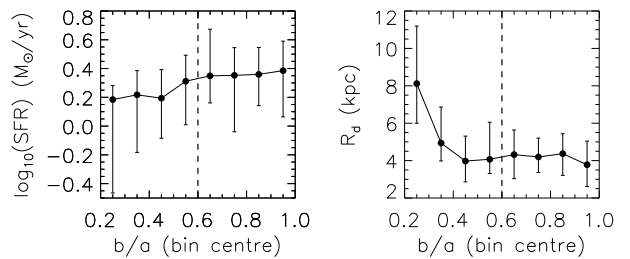}
	\caption{Medians and 1$\sigma$ intervals of SFR (left panel) and $R_d$ (right panel) for bulge MWAs binned by b/a, in the case where no b/a cut is applied to the parent sample. The dashed lines mark where b/a $= 0.6$, below which galaxies are \textit{not} included in the parent sample and so not included in any subsequent analysis} 
	\label{bafig}
	\end{center}
\end{figure}

Measured bulge MWA properties are relatively sensitive to viewing angle, given the status of bulge MWAs as disc dominated systems. In \autoref{bafig}, we the SFR and $R_d$ values of bulge MWAs in bins of b/a, in the case where no b/a cut is applied to the parent sample. We find a small but non-negligible drop in SFR at b/a $< 0.6$ along with a significant increase in $R_d$ at b/a values below 0.4. Thus, a b/a cut is necessary to avoid biases in the properties of sample galaxies.

We select  \textbf{``scale MWAs"} based on stellar mass and on exponential disc scale length. The MW scale length has repeatedly been suggested to be atypically short for the MW's stellar mass; thus, scale length is potentially important for understanding the MW's place in the wider extragalactic context \citep{boardman2020}. We use the same mass cut as before, and we further cut the sample to include only galaxies with $R_d$ values between 2.51 kpc and 2.93 kpc; this is based on the 1-$\sigma$ intervals reported in LN16 from optical data only, as opposed to the intervals from IR or optical+IR data.

\citet{licquia2016a} discuss at length the applicability of the MW scale length for comparing with external galaxy measurements in their Section 5.2.1; we provide a brief summary here, and direct the interested reader to that paper for a more complete discussion. MW disc scale length calculations are typically carried out via star-count analyses, which differs significantly from the photometric methods employed for other galaxies. On the other hand, the ratio between the MW's IR and visible scale lengths appears similar to that measured for external galaxies, and MW dynamical scale length measurements are consistent with measurements made through other means (LN16); thus, the MW scale lengths reported in the literature appear robust for our purposes.

Lastly, we select a sample of \textbf{``morphological MWAs"}, defined as those galaxies with masses in the 1$\sigma$ MW range that also possess bar and spiral features. The presence of a bar in particular is likely important in understanding a given galaxy's properties \citep[e.g.][]{dk2020}, making it worthwhile to explore barred spiral analogs separately from the previously-defined samples. We largely follow Table 3 of \citet{willett2013} in performing this selection: we restrict to galaxies satisfying \textit{smooth} $\leq 0.57$, \textit{edgeon} $\leq 0.285$, and $N_{bar,spiral} \geq 20$. For conservative sample selections, \citet{willett2013} further suggest using minimum vote thresholds of $0.8$ for selecting on morphological features, and we employ those thresholds on the \textit{bar} and \textit{spiral} vote fractions. We show some examples morphological MWAs in \autoref{morphmwaexamples}. 

\begin{figure}
\begin{center}
	\includegraphics[trim = 1cm 18.cm 0cm 6.5cm,scale=1.1,clip]{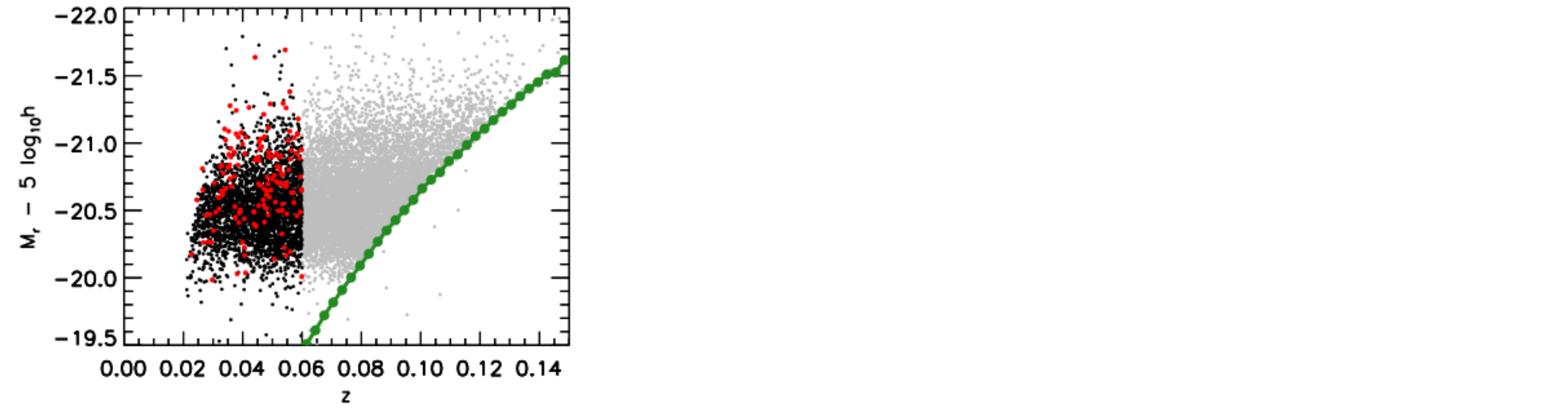}
	\caption{Plot of r-band absolute magnitude vs redshift for the bulge MWA sample (red points), along with all galaxies satisfying the MW mass cut inside (black points) and outside (grey points) the volume-limited redshift region. The green points show the 99th percentile absolute magnitude of the parent sample as a function of redshift.} 
	\label{mwas_zmag}
	\end{center}
\end{figure}

We restrict all four MWA samples to galaxies of redshifts $z \leq 0.06$, in order to ensure that the samples are volume-limited. In \autoref{mwas_zmag}, we show the magnitudes and redshifts of the bulge MWA sample along with all galaxies that satisfy the MW mass cut. By limiting to $z \leq 0.06$, we obtain volume-limited samples as desired; the same situation occurs for the other three MWA samples.

\begin{figure*}
\begin{center}
	\includegraphics[trim = 1cm 2.5cm 1cm 3.5cm,scale=0.7,clip]{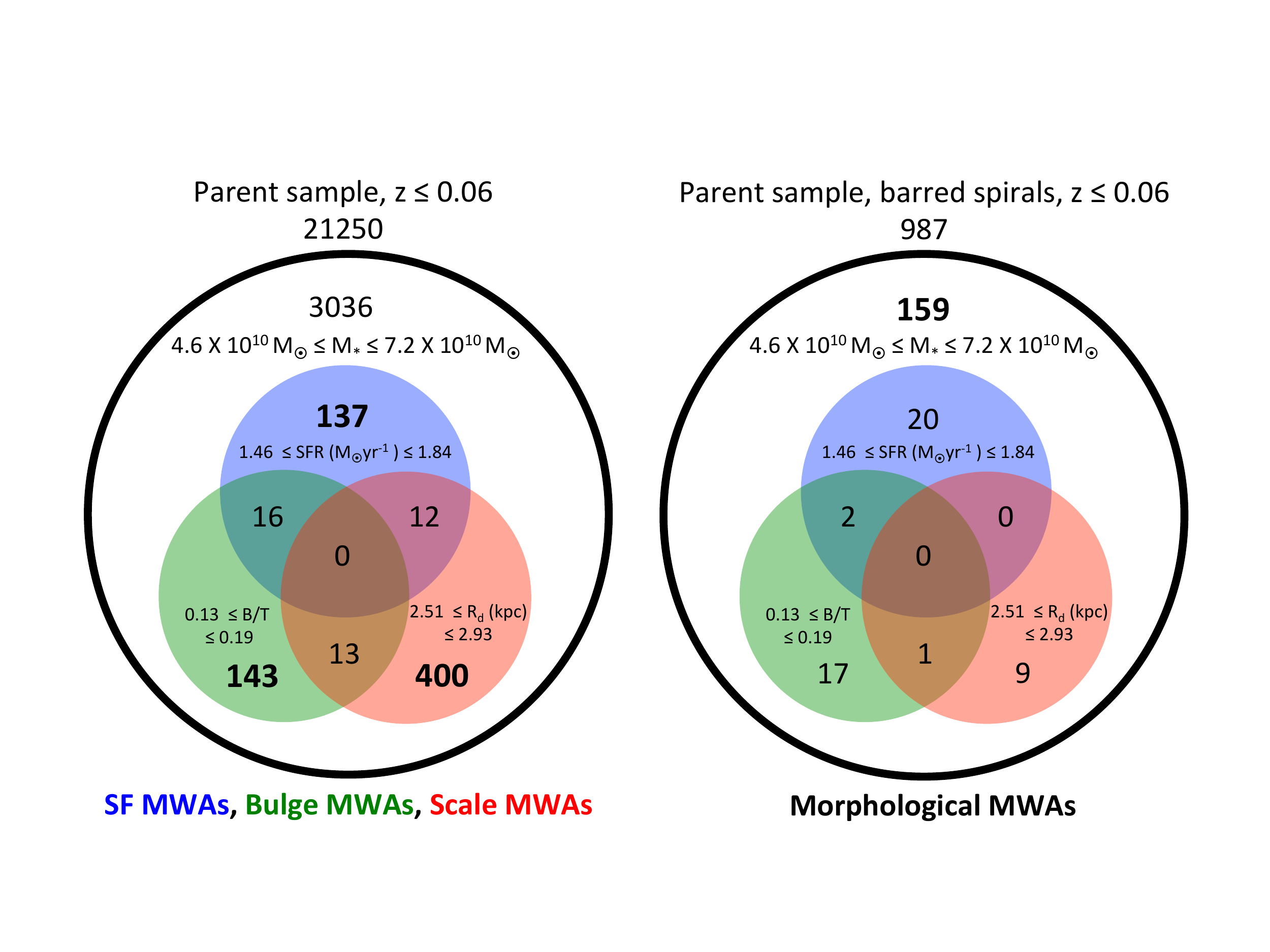}
	\caption{Venn diagrams showing the numbers of galaxies satisfying the MWA cuts discussed in the text. Barred spirals are defined by Galaxy Zoo vote fractions as described in the text for the morphological MWAs.}
	\label{mwavenn}
	\end{center}
\end{figure*}

In \autoref{mwavenn}, we show the size of the selected MWA samples in the form of a Venn diagram, along with the numbers of galaxies satisfying different combinations of parameter cuts. We find that the number of identified MWAs becomes vanishingly small as the number of criteria is increased, with not one MWA satisfying all of the cuts. This problem of dimensionality was previously highlighted in \citet{frasermckelvie2019b}, and demonstrates the need to employ just a few selection criteria when large MWA samples are required.

Many additional possible "analog" definitions exist beyond the ones we consider here. The nearby NGC 891, for instance, is often considered an MWA on the basis of morphology and rotational velocity \citep[e.g.][]{mouhcine2010,hughes2014}. \citet{kormendy2019} argue NGC 4565 and NGC 5746 to be analogs on the basis of their morphology, and in particular on the presence of boxy pseudobulges. A search for boxy bulge structures is possible with SDSS imaging \citep{yoshino2015} but would require edge-on samples, and so conflicts with the requirements of the other analog samples in this work. An analog sample based on rotational velocities is more feasible but of questionable additional value, particularly in light of the stellar mass Tully-Fisher relation \citep[TFR;][]{tully1977,mocz2012,licquia2016a}. It should also be noted that in \citet{licquia2016a}, the quoted uncertainty in the MW rotational velocity is larger than the uncertainty in the stellar mass, relative to the scatter in the stellar mass TFR itself.

\subsection{M31 analog sample selections}\label{samplem31as}

As with the MW, a range of calculations of key M31 properties have been performed over the years. M31 is known to be more massive than the MW, and has consistently been measured to possess a significantly larger disc scale length \citep[e.g.][]{hammer2007}. In addition, M31 is generally agreed to possess a lower current star-formation rate than the MW \citep[e.g.][and references therein]{yin2009}. Our M31 sample selections are designed to capture this behaviour, while allowing for the spread in reported M31 measurements. We summarise our M31A selections in \autoref{m31asampletable}, and we explain our selections over the remainder of this subsection.

\begin{table*}
\begin{center}
\begin{tabular}{l|c|c|c|c}
Analog Sample & SFR ($M_\odot$~yr$^{-1}$) & B/T & $R_d$ (kpc) & Galaxy Zoo vote fraction \\
\hline
\hline
M31 star-formation analogs & $0.41-0.83$ & --- & --- & --- \\
M31 bulge analogs & --- & $0.24-0.42$ & --- & --- \\
M31 scale analogs & --- & --- & $4.8-5.8$ & --- \\
M31 morphological analogs & --- & --- & --- &  spiral $> 0.8$, bar $<0.2$, $N_{\rm spiral} \geq 20$\\
 &  &  & &  smooth $\leq 0.57$, edgeon $\leq 0.285$ \\
\end{tabular}
\end{center}
\caption{M31A sample definitions, in addition to $M_* = 9.9 \times 10^{10} - 1.09 \times 10^{11} M_\odot$, $z \leq 0.09$, $P_{pS} \leq 0.32$ and $b/a \geq 0.6$.}
\label{m31asampletable}
\end{table*}

Firstly, we restrict all M31A selections to galaxies with masses between $9.9 \times 10^{10} M_{\odot}$ and $1.09 \times 10^{11} M_{\odot}$. We obtained this range from \citet{mutch2011}, who computed it from from the semi-analytic mass modelling of \citet{geehan2006} and other compiled literature values \cite[e.g.][]{barmby2006,hammer2007}.

We select a sample of \textbf{``star-forming M31As"} (hereafter \textbf{``SF M31As")} using the M31 SFRs calculated over 10 Myr by \citet{kang2009} , as presented in their Table 3 based on combined UV and IR photometry. Taking for boundaries the values calculated with stellar model grids with subsolar ($Z = 0.008$) and supersolar ($Z = 0.05$) metallicities respectively, we obtain an SFR range of 0.41-0.83 $M_{\odot}/yr$. This is broadly consistent with the range of values reported in past literature \citep[see for instance Table 1 of][]{yin2009}. Where necessary, we assume a midpoint M31 SFR of 0.46 $M_{\odot}/\textrm{yr}$, calculated with solar-metallicity ($Z = 0.02$) model grids.

We select \textbf{``bulge M31As"} by considering all results contained within Table 5 of \citet{tamm2012}, along with the results of \citet{barmby2006} and \citet{courteau2011}. We take the mean and standard deviation of the resulting B/T values, obtaining a value of $B/T_{M31} = 0.33 \pm 0.09$; we use this value to define our M31 structural analog sample, employing a cut of $B/T = 0.24 - 0.42$

We select a sample of \textbf{``scale M31As"} using the $5.3 \pm 0.5$ kpc M31 disc scale length reported by \citet{courteau2011} from infra-red \textit{Spitzer}/IRAC imaging. This is slightly lower than the majority of photometric measurements, as can be seen in \citet{hammer2007} and \citet{yin2009} along with references therein; \citet{hammer2007} use a value of $5.8 \pm 0.4$ kpc to represent the range of reported photometric  values, for instance. However, the \citet{courteau2011} value is more analogous to the (effectively mass-weighted) value we employ for the MW and less sensitive to dust extinction effects than measurements made in bluer bands. In addition, the choice between the M31 $R_d$ values of \citet{courteau2011} and \citet{hammer2007} matters little in practice, as they are both consistent with M31 having a scale length that is reasonably standard for a galaxy of its mass \citep[e.g.][]{fathi2010,wu2018}. 

\begin{figure}
\begin{center}
	\includegraphics[trim = 5.6cm 3cm 10cm 2cm,scale=0.9
	,clip]{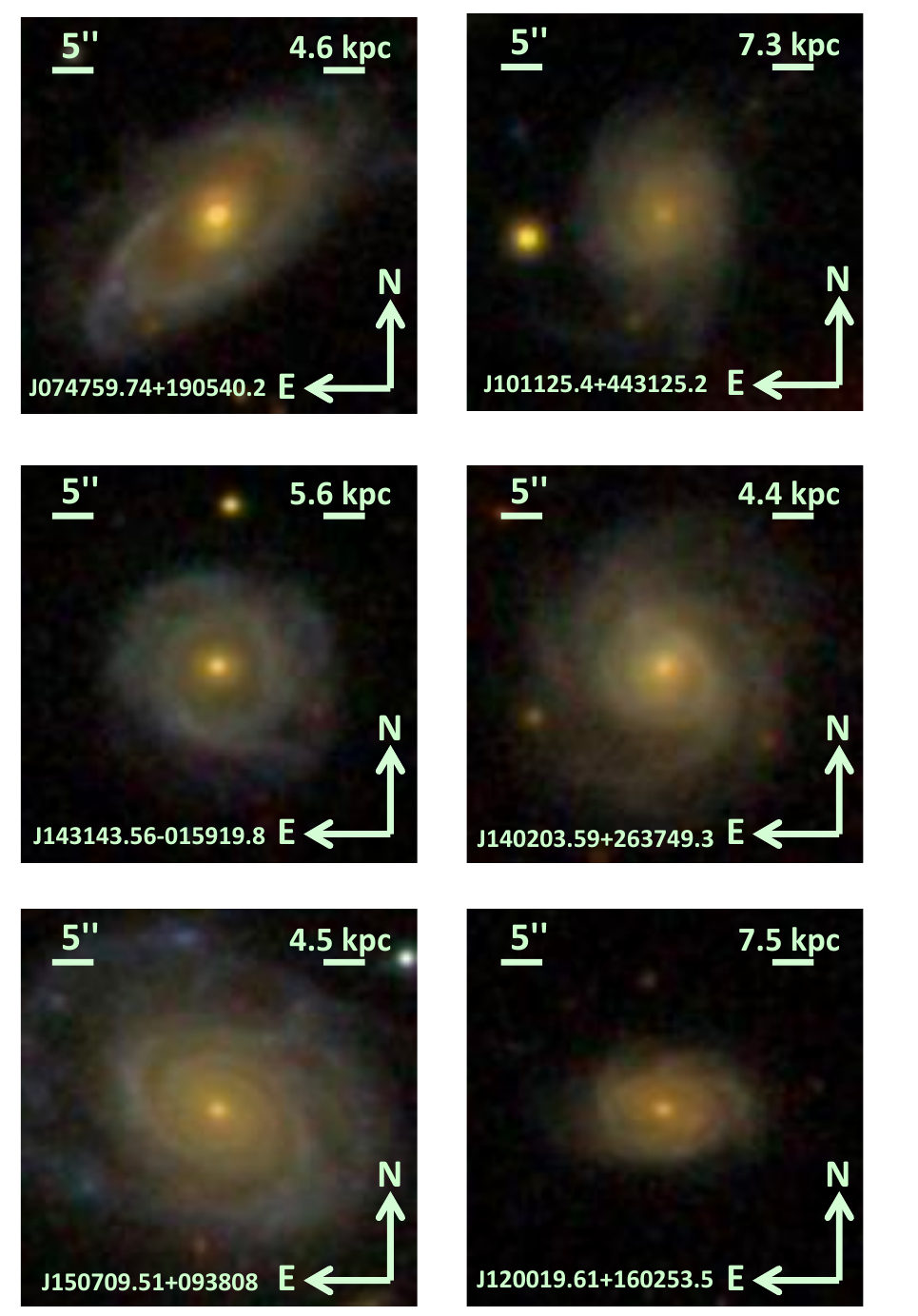}
	\caption{Example morphological M31As, selected as discussed in the text.} 
	\label{morphm31aexamples}
	\end{center}
\end{figure}

Finally, we select a sample of \textbf{``morphological M31As"} based on  Galaxy Zoo 2 vote fractions. M31 is known to contain a stellar bar \citep[e.g.][]{beaton2007}, but visually appears as an unbarred spiral galaxy \citep[e.g.][]{sandage1981}; thus, we select for the presence of spiral features along with the apparent \textit{absence} of bar features. As for the MWAs, we restrict to galaxies satisfying \textit{smooth} $\leq 0.57$, \textit{edgeon} $\leq 0.285$. We then use selection thresholds of spiral $> 0.8$ and bar$< 0.2$, along with requiring $N_{spiral} \geq 20$; we impose no requirement on $N_{bar}$ in this case, as low \textit{bar} values typically correspond to low $N_{bar}$ numbers. We present some example morphological M31As in \autoref{morphm31aexamples}. 

\begin{figure}
\begin{center}
	\includegraphics[trim = 1cm 17.5cm 0cm 5.3cm,scale=1.1]{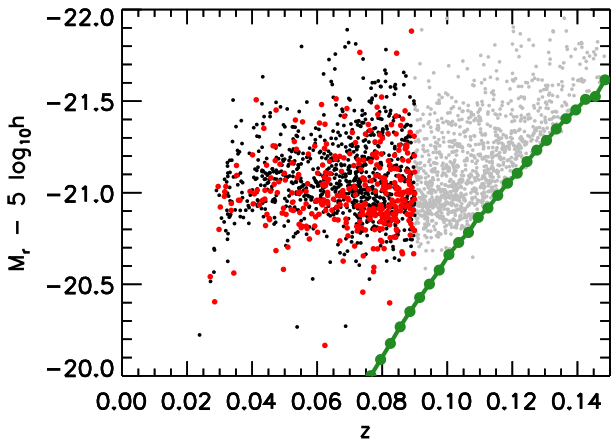}
	\caption{Plot of r-band absolute magnitude vs redshift for the bulge M31A sample (red points), along with all galaxies satisfying the M31 mass cut inside (black points) and outside (grey points) the volume-limited redshift region. The green points show the 99th percentile absolute magnitude of the parent sample as a function of redshift.} 
	\label{m31as_zmag}
	\end{center}
\end{figure}

\begin{figure*}
\begin{center}
	\includegraphics[trim = 1cm 2.5cm 1cm 3.5cm,scale=0.7,clip]{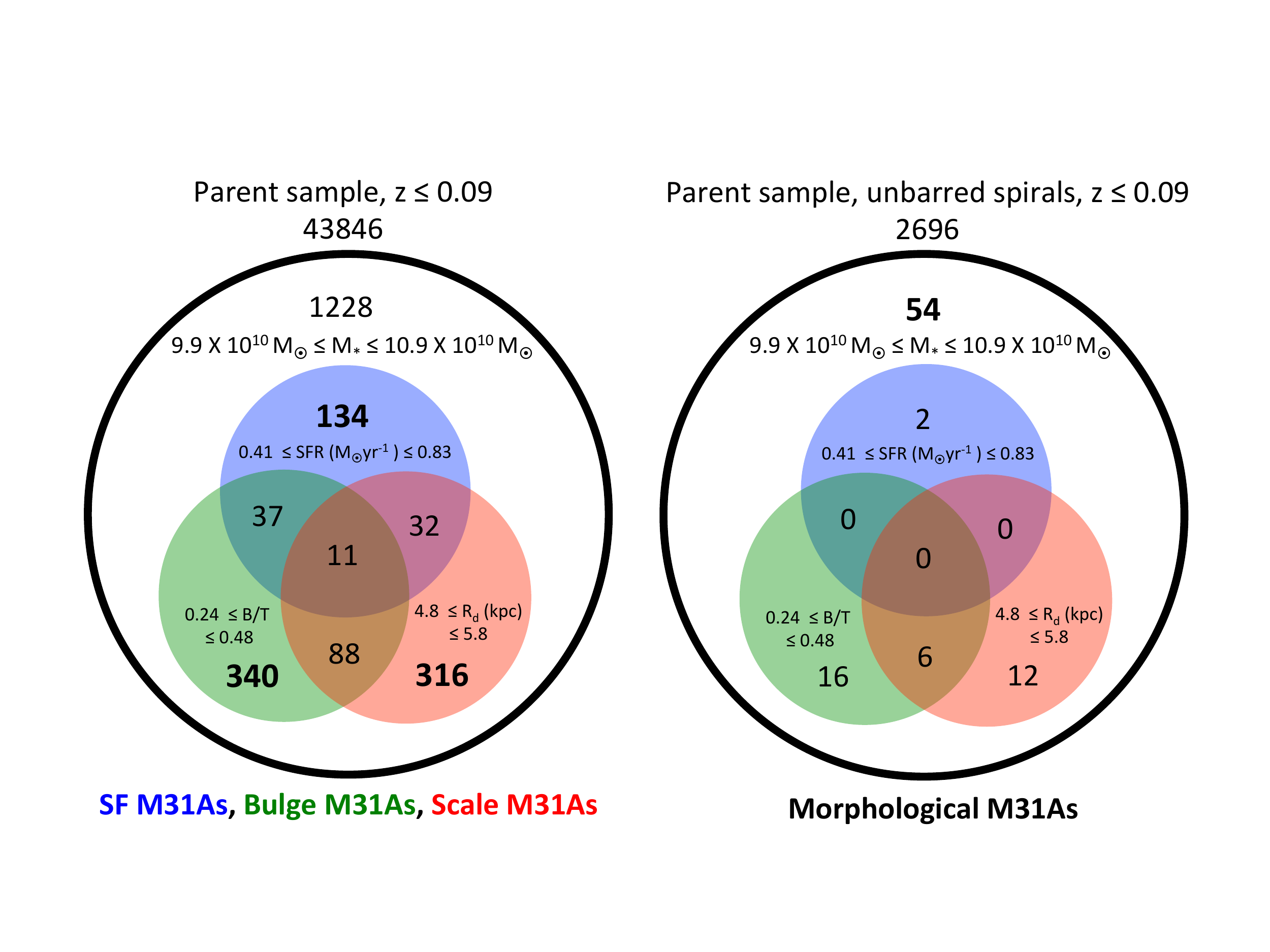}
	\caption{Venn diagrams showing the numbers of galaxies satisfying the M31A cuts discussed in the text. Unbarred spirals are defined by Galaxy Zoo vote fractions as described in the text for the morphological M31As} 
	\label{m31avenn}
	\end{center}
\end{figure*}

We restrict all M31A samples to galaxies with redshifts $z \leq 0.09$ to ensure volume-limiting; we are able to use a higher maximum redshift in this case due to the brighter magnitudes displayed by the M31As. We present the ranges of redshifts and magnitudes of the bulge M31A samples in \autoref{m31as_zmag}, wherein we show that our adopted redshift cut leads to a volume-limited sample as desired; the same situation is seen in all M31A samples. We present Venn diagrams detailing the M31A sample sizes in \autoref{m31avenn}.

\section{Results}\label{samplecompare}

\subsection{MWA results}\label{mwasamplecompare}

\begin{figure*}
\begin{center}
	\includegraphics[trim = 1.5cm 7.5cm 0.5cm 15cm,scale=1.1]{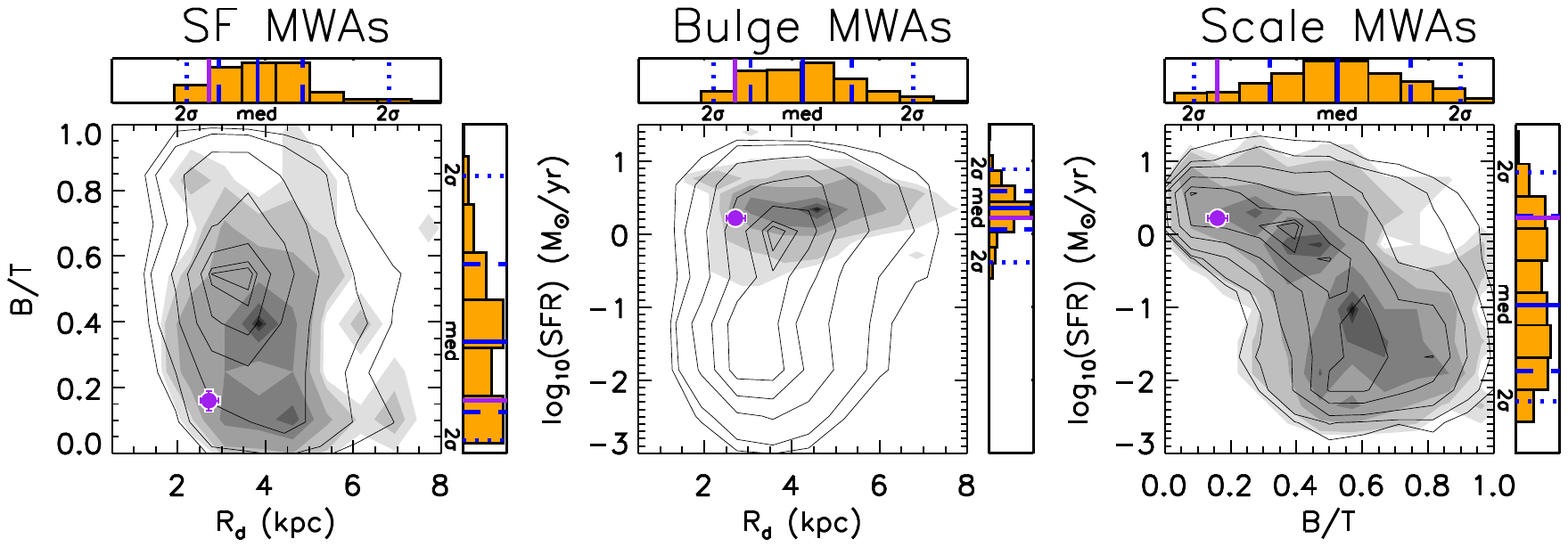}
	\caption{2D and 1D histograms of parameter distributions for SF MWAs, bulge MWAs and scale MWAs, with the MW values shown in purple. The grey-scale surfaces represent the specific MWA samples, while the contours represent all galaxies that satisfy the MW mass cut. The contour levels represent 5\%, 10\%, 25\%, 50\%, 75\%, 90\% and 95\% of the largest single bin. The blue solid lines represent the medians in a given parameter, the dashed lines the 1$\sigma$ ranges, and the dotted lines the 2$\sigma$ ranges.} 
	\label{mwas2dhist}
	\end{center}
\end{figure*}

In the left panel of  \autoref{mwas2dhist}, we show the B/T and $R_d$ distributions of the SF MWAs as a 2D histogram, along with the MW value and contours of properties for all galaxies that satisfy the MW mass cut. We find the MW to lie within 1$\sigma$ of the B/T values calculated for SF analogs, though we note that the range of B/T values amongst SF MWAs is wide. The MW is somewhat of an outlier in terms of scale length, but remains within the 2$\sigma$ region.

We plot the SFR and $R_d$ values of the bulge MWAs in the middle panel of \autoref{mwas2dhist}, in which we find the adopted MW SFR to lie well within $1\sigma$ of the bulge MWAs. We find the MW disc scale length to again be low compared to the majority of bulge MWAs, though still within 2$\sigma$ of this sample.

\begin{figure*}
\begin{center}
	\includegraphics[trim = 0.5cm 18cm 1cm 4cm,scale=0.9]{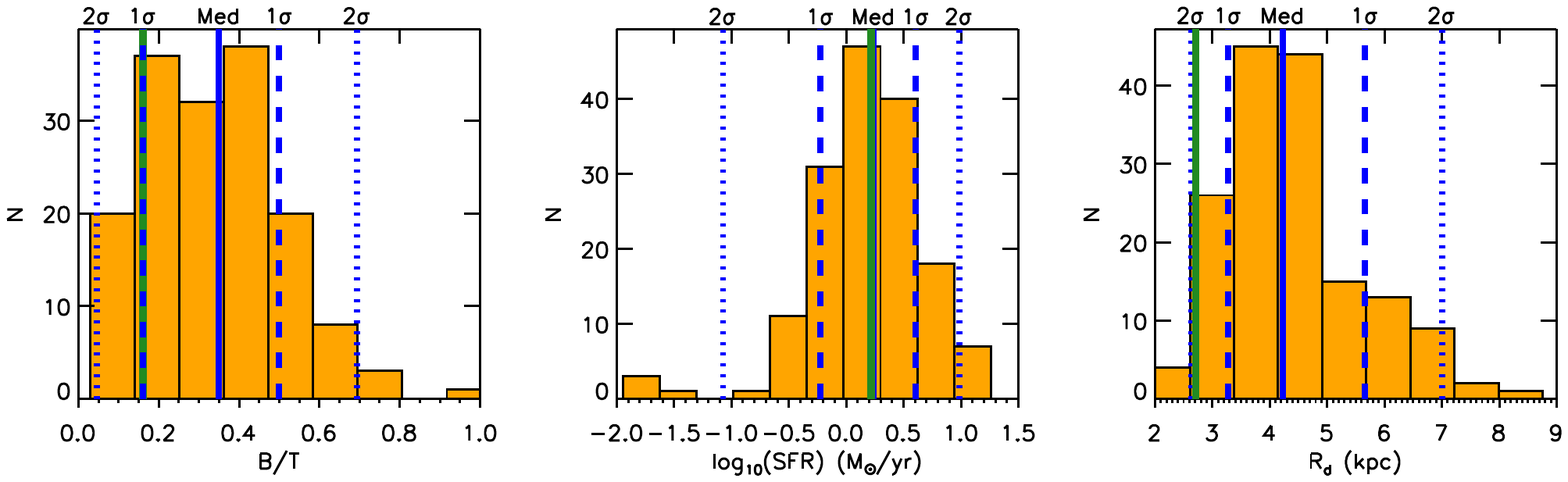}
	\caption{1D histograms of parameter distributions for morphological MWAs. The blue solid lines show the medians of the MWA samples; the dashed blue lines indicate the 1$\sigma$ regions, and the dotted blue lines indicate the 2$\sigma$ regions. The central MW value is also shown in each window as a solid green line.} 
	\label{mbar2dhist}
	\end{center}
\end{figure*}

The right panel of \autoref{mwas2dhist} presents the properties of the scale MWAs in terms of SFR and B/T, following the same format as the preceding two panels. We see the expected star-forming/quiescent galaxy dichotomy in the mass-selected sample, with a discy star-forming component along with a quiescent and more bulgey component; we find this dichotomy to be maintained amongst the scale analogs, albeit weighted more heavily towards the quiescent component. We also find amongst the discy scale analogs, the MW \textit{does not appear to be unusual}, with both an SFR and B/T in good consistency with the behaviour of the discy scale analogs. 

We present in \autoref{mbar2dhist} 1D histograms of properties of the morphological MWAs in terms of SFR, B/T and $R_d$ in turn. We caution though that the B/T and $R_d$ values are specifically calculated from two-component photometric fits, which will not fully capture the photometry of barred galaxies and can significantly overestimate barred galaxies' bulge-to-total ratios \citep{laurikainen2005,laurikainen2006,kruk2018}; thus, we include B/T and $R_d$ values purely for completeness in this case. In terms of SFR, we find  the MW to well within $1\sigma$ of the morphological MWA sample.

Overall, the SF, bulge and morphological MWAs are broadly consistent in their range of properties, with all three favouring the selection of disc-dominated star-forming galaxies. It is apparent that MWA samples do \textit{not} accurately predict the MW's short scale length unless the scale length is specifically selected on, and it is also apparent that the scale MWA sample is not particularly constraining in itself. At the same time, the MW appears to not be the least bit unusual amongst the star-forming galaxies within the scale MWA sample. From these findings, we argue that scale length should be considered \textit{in addition} to other parameters in order to select the most stringent MWA galaxies.

\subsection{M31A results}\label{m31asamplecompare}

\begin{figure*}
\begin{center}
	\includegraphics[trim = 1.5cm 7.5cm 0.5cm 15cm,scale=1.1]{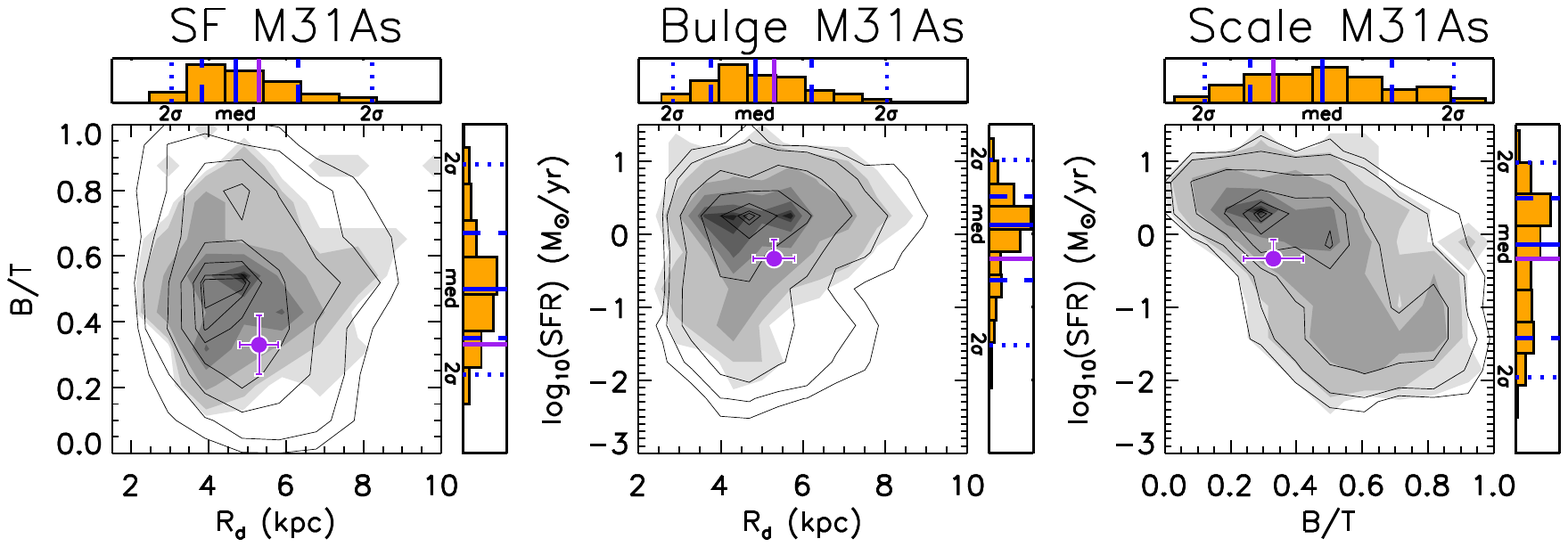}
	\caption{2D and 1D histograms of parameter distributions for SF M31As, bulge M31As and scale M31As, with the M31 values shown in purple. The grey-scale surfaces represent the specific M31A samples, while the contours represent all galaxies that satisfy the M31 mass cut. The contour levels represent 5\%, 10\%, 25\%, 50\%, 75\%, 90\% and 95\% of the largest single bin. The blue solid lines represent the medians in a given parameter, the dashed lines the 1$\sigma$ ranges, and the dotted lines the 2$\sigma$ ranges.} 
	\label{m31apar2dhist}
	\end{center}
\end{figure*}

We plot the SF M31A sample in terms of B/T and $R_d$ in the left panel of \autoref{m31apar2dhist}, with our adopted M31 values shown on the same figure along with contours of all galaxies satisfying our M31 mass cut. We find M31's scale length to be in excellent agreement with the SF M31As. However, we find the SF M31As to over-predict the B/T to an extent, with M31's B/T offset low by approximately $1\sigma$.

We present in the middle panel of \autoref{m31apar2dhist} the bulge M31A sample in terms of SFR and $R_d$. We find the M31 SFR to be slightly low amongst the bulge M31As, though still well within 1$\sigma$ of this sample's range. We also find the scale lengths of the bulge M31As to agree excellently with that of M31.=

We present the SFR and B/T distributions of the scale M31A sample in the right panel of \autoref{m31apar2dhist}. We find M31 to within $1\sigma$ in terms of both properties when the properties are considered individually, but it is clear that M31's SFR is somewhat lower than is typical for scale M31As in M31's likely B/T range.

\begin{figure*}
\begin{center}
	\includegraphics[trim = 0.5cm 18cm 1cm 4.5cm,scale=0.9]{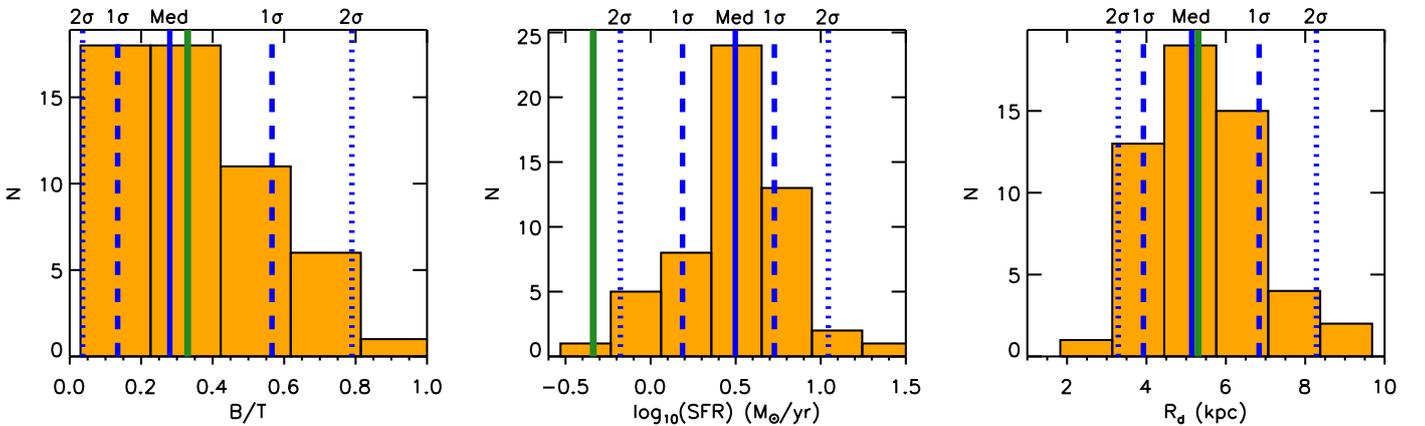}
	\caption{1D histograms of parameter distributions for morphological M31As. The blue solid lines show the medians of the M31As; the dashed blue lines indicate the 1$\sigma$ regions, and the dotted blue lines indicate the 2$\sigma$ regions. The central M31 value is also shown in each window as a solid green line.} 
	\label{morphm31a2dhist}
	\end{center}
\end{figure*}

Lastly, we present the properties of the morphological M31As in \autoref{morphm31a2dhist}. We find our adopted central M31 SFR to be offset low by more than 2$\sigma$ in this case, while finding M31's B/T and $R_d$ to both not be the least bit unusual amongst the morphological M31As. 

To summarise, we find the scale length of M31 to agree excellently with those of the M31A samples, but we find the SFR of M31 to be somewhat low. Although M31's SFR is in reasonable consistency with the bulge and scale M31A samples, it is significantly offset from the morphological M31As. In addition, M31's low SFR leads the SF M31A sample to over-predict M31's B/T by around $1\sigma$. From this, we advocate including mass and B/T in selection criteria for targeting the closest M31As, with M31's low SFR also important to keep in mind.

\section{Discussion and conclusion}\label{discussion}

In this paper, we have explored the properties of various galaxy samples selected as ``Milky Way Analogs" or ``Andromeda Analogs" using various selection criteria. We compared the properties of both the MW and M31 to their respective analogs, with the aim of understanding how to best select constraining ``analog" samples. 

Critically, the MWA samples do \textit{not} accurately predict the short disc scale length of the MW unless the scale length is included in the selection criteria. Although a few reported MW $R_d$ values are highly consistent with our non-scale MWA samples \citep[e.g.][]{yy1992,benjamin2005,chang2011,grady2020}, the vast majority of measurements are less so \citep[e.g.][]{kent1991,rm1991,chen1999,bovy2013,mao2015,li2018a}. The scale length of the MW disc is therefore important to consider if one wishes to select the most similar MWA galaxies.

It should be noted that the MW’s scale length is \textit{not unduly unusual} amongst those of the MW’s direct peers, with the MW scale length remaining within 2$\sigma$ of the SF and bulge MWAs. In turn, the MW appears not the least bit unusual amongst star-forming scale MWAs, with the MW's SFR and B/T values both well within the scale MWA sample's range. Thus, while the MW is indeed an outlier in terms of $R_d$, it is not an  outlier to the extent of requiring a unique explanation.

Aside from the MW's short scale length, the MW is likewise not unusual amongst the other MWA samples considered. In terms of B/T, the MW falls within the 1$\sigma$ range displayed by the SF MWAs; likewise, the SFR of the MW falls well within 1$\sigma$ of the SFR range of the bulge MWAs. We also find the MW SFR to be well within $1\sigma$ of the morphological MWAs. Overall, we find the MW to be a relatively typical galaxy aside from its short disc scale length, consistent with previous work \citep[e.g.][]{hammer2007, bovy2013,licquia2016a}.

In contrast to the MW, we find M31's SFR to be over-predicted to various extents in all non-SF M31A samples. In turn, the SF M31As over-predict M31's B/T. Thus, we argue that the combination of a low SFR and relatively low B/T are important in selecting the closest M31A galaxies. Our results here are in good consistency with the work of \citet{mutch2011}, who argue M31's properties to be consistent with galaxies on the green valley region of the color-magnitude diagram. Compared to the MW, we also find M31's disc scale length to be in much greater consistency with its respective analog samples, which is likewise consistent with previous work \citep[e.g.][]{hammer2007}.

Similar results can be expected if one parametrises galaxy mass with $V_{rot}$ instead of stellar mass, due to the existence of the stellar mass TFR. We confirmed this by cross-matching our parent sample with the ALFALFA extragalactic \HI\ source catalog \citep{haynes2018}, converting velocity widths to $V_{rot}$ using inclinations obtained from \citet{simard2011}. We adopted MW and M31 $V_{rot}$ values of 220 $\pm$ 22~km~s$^{-1}$ \citep{kerr1986} and 226 $\pm$ 29~km~s$^{-1}$ \citep{carignan2006}, respectively, where we have adopted 10\% errors for the MW in line with previous work \citep[e.g.][]{licquia2016a}; this leads to 1$\sigma$ $V_{rot}$ windows of 198-242~km~s$^{-1}$ and 197-255~km~s$^{-1}$ for the MW and M31 in turn. Due to the resulting overlap in the MW and M31 values, we selected a single analog sample to cover $V_{rot}$ values between 197 and 255~km~s$^{-1}$; we restricted this selection to galaxies with inclinations no lower than $40^o$ and with velocity width errors no greater than 15~km~s$^{-1}$. The resulting sample was found to be essentially complete, so we made no explicit redshift cut. We found the MW scale length to be offset low by over $1\sigma$ in this case, with our adopted central M31 SFR also low by roughly $1\sigma$; both galaxies are otherwise in good consistency with this analog sample. Our results from this sample are therefore similar to what we find with the other MWA and M31A samples. However, the use of \HI\ data strongly biases the cross-matched parent sample towards disc-dominated star-forming galaxies, and we prefer our other analog selections for this reason.

The difference in MWA and M31A results - particularly in terms of scale lengths - likely relate to the respective evolution histories of the MW and M31. \citet{mackereth2018} report that bimodal alpha abundance ratios only appear in EAGLE galaxies when those galaxies have had a particularly violent early merging history, though other simulation works \citep[e.g.][]{grand2018,buck2020} provide counterpoints. That the MW experienced an early merging history is supported by the age-metallicity distribution of MW globular clusters \citep{kruijssen2019} and by substructures apparent in Gaia data \citep{helmi2018,belokurov2018,elias2020}, and the apparent short scale length of the MW can also be understood on this basis \citep{mo1998}. M31 likely had a more extended merging history, meanwhile, that is more typical of spiral galaxies \citep{hammer2007}. Such a notion is consistent with the apparent burst in star-formation that occurred in M31 roughly 2-4 Gyr ago \citep{williams2015}. This, along with the comparatively large amount of disc heating evident in M31 \citep{hammer2018} and the multiple substructures around M31 with similar stellar populations \citep{bernard2015}, supports the idea of major merger occurring at around that time, with M32 being a possible remnant of the merging galaxy \citep{dsouza2018}.

Looking to larger galaxy samples, it has been reported that smaller galaxies display stellar populations that are on average older and more metal-rich at a given galaxy mass \citep[e.g.][]{scott2017,zheng2018}. This is likewise consistent with the scenario of the MW having an atypical formation history amongst spiral galaxies, and further highlights the importance of considering disc scale length when setting out to select the most similar MWAs. 

Given the apparent importance of scale length in understanding the MW, we further argue that scale length should be specifically considered when attempting to reproduce the MW in models and simulations, along with properties such as B/T and SFR. Simulation studies often focus on galaxies of MW-like mass (halo, stellar or total), with environment also frequently taken into account \citep[e.g.][]{scannapieco2015,nuza2019,carlesi2020,santistevan2020}. Additional selections can be imposed to ensure disc-domination in simulated galaxies identified as being MW-like, as for instance done in \citet{mackereth2018}, but such selections can still be expected to yield galaxies that are much more extended than the MW on average. By considering in simulations galaxies that match the MW in terms of a wider range of properties - in particular, by considering galaxies that are star-forming and disc-dominated while also possessing compact discs - the opportunity exists to further understand how our own galaxy came to be.

To summarise, we find the MW's properties to mostly be in good consistency with its analogs, irrespective of the particular selection strategy employed. We do however find the MW disc to be atypically compact compared to the MWA samples, meaning that MWA samples do \textit{not} accurately predict the MW's disc scale length unless the scale length is included in the selection criteria. We therefore argue that scale length should be considered \textit{in addition} to other parameters when selecting the most stringent MWAs. We find M31's scale length to be in excellent agreement with its analogs, though its SFR is evidently lower than the majority of its structural peers. Thus, we advocate selecting on a combination of mass, SFR and bulge fraction to choose the closest M31As.

\section*{Data Availability}

The data underlying this article are available in its online supplementary material. We include as online supplementary material two tables detailing all MWAs and M31As, with no redshift cut applied. We show the first five rows of each of these tables in \autoref{mwatab} and \autoref{m31atab}.

\begin{table*}
\begin{center}
\begin{tabular}{c|c|c|c|c|c|c|c|c|c|c}
Objid (SDSS DR8) & RA & DEC & z & $\log_{10}(M_*$ & $\log_{10}(\mathrm{SFR}$ & B/T & $R_d$ (kpc) & \textit{spiral} & \textit{bar} & Flag \\
 & (deg) & (deg) & & $/M_\odot)$  &$/M_\odot yr^{-1})$ & & & & & \\
\hline
\hline
1237663917872054658 & 111.586 & 43.533  & 0.057 & 10.75 & -2.21 & 0.40 & 2.75 & 0.00 & 0.00 & 0\\
1237663547431518682 & 111.638 & 37.914  &  0.082 & 10.79 & -1.48 & 0.49 & 2.88 & 1.00 & 0.98 & 0\\
1237663916797723168 & 111.981 & 41.959  & 0.058 & 10.81 & 0.18 & 0.41 & 4.36 & 0.61 & 0.93 & 0\\
1237663917335052608 & 112.048 & 42.981  & 0.066 & 10.76 & 0.41 & 0.81 & 4.02 & 0.97 & 1.00 & 1\\
1237663547432108541 & 112.500 & 39.100 & 0.089 & 10.84 & 0.23 & 0.17 & 6.05 & 0.09 & 0.98 & 0\\
\end{tabular}
\end{center}
\caption{Table of Milky Way Analogs, with no redshift cut applied. A flag value of "1" indicates a galaxy's status as a morphological MWA. We show the first five rows here; the full table will be made available online.}
\label{mwatab}
\end{table*}

\begin{table*}
\begin{center}
\begin{tabular}{c|c|c|c|c|c|c|c|c|c|c}
Objid (SDSS DR8) & RA & DEC & z & $\log_{10}(M_*$ & $\log_{10}(\mathrm{SFR}$ & B/T & $R_d$ (kpc) & \textit{spiral} & \textit{bar} & Flag \\
 & (deg) & (deg) & & $/M_\odot)$  &$/M_\odot yr^{-1})$ & & & & & \\
\hline
\hline
1237663917872185769 & 111.824 & 43.785 & 0.057 & 11.00 & 0.40 & 0.45 & 5.28 & 0.17 & 0.65 & 0\\
1237663916797985250 & 112.351 & 42.429 & 0.133 & 11.02 & 1.11 & 0.18 & 5.75 & 0.72 & 1.00 & 0\\
1237663530252238968 & 112.639 & 39.049 & 0.088 & 11.01 & 0.39 & 0.27 & 5.77 & 0.89 & 1.00 & 0\\
1237663787414782244 & 114.002 & 44.470 & 0.079 & 11.01 & -0.88 & 0.62 & 5.77 & 0.00 & 1.00 & 0\\
1237657594607501971 & 114.303 & 27.235 & 0.093 & 11.01 & 0.27 & 0.42 & 5.49 & 0.00 & 0.97 & 0\\
\end{tabular}
\end{center}
\caption{Table of Andromeda Analogs, with no redshift cut applied. A flag value of "1" indicates a galaxy's status as a morphological M31A. We show the first five rows here; the full table will be made available online.}
\label{m31atab}
\end{table*}

\section*{Acknowledgements}

We thank the anonymous referee for their comments, which served to greatly improve the clarity of this paper. We thank Cristina Chiappini for her useful and insightful comments on this manuscript. 

The support and resources from the Center for High Performance Computing at the University of Utah are gratefully acknowledged.

\bibliographystyle{mnras}
\bibliography{bibliography}

\bsp	
\label{lastpage}
\end{document}